\documentclass[11pt]{pazh}

\usepackage{psfig}
\usepackage{amssymb}

\begin{document}
        
\setcounter{page}{25}

\sloppypar

\title{\bf UGC\,7388: a galaxy with two tidal loops}

\author{M. Fa\'undez-Abans\inst{1}, V.P. Reshetnikov\inst{2}, 
M. de Oliveira-Abans\inst{1}, and I.F. Fernandez\inst{1}} 

\institute{MCT/Laborat\'{o}rio Nacional de Astrof\'{\i}sica, Itajub\'{a}, Brasil
\and
St.Petersburg State University,
Universitetskii pr. 28, Petrodvoretz, 198504 Russia
}

\authorrunning{Fa\'undez-Abans}
\titlerunning{UGC\,7388}

\abstract{We present the results of spectroscopic and morphological 
studies of the galaxy UGC\,7388 with the 8.1-m Gemini North
telescope. Judging by its observed characteristics, UGC\,7388 is a giant 
late-type spiral galaxy seen almost edge-on. The main body of the galaxy is 
surrounded by two faint ($\mu_{B} \sim 24^m/\Box\arcsec$ and 
$\mu_{B} \sim 25.^m5/\Box\arcsec$) extended ($\sim$20--30 kpc) loop-like structures. 
A large-scale rotation of the brighter loop about the main galaxy has been detected.
We discuss the assumption that the tidal disruption of a relatively
massive companion is observed in the case of UGC\,7388. A detailed study 
and modeling of the observed structure of this unique galaxy can give important 
information about the influence of the absorption of massive companions on the 
galactic disks and about the structure of the dark halo around UGC\,7388.
\keywords{galaxies, interacting galaxies, morphology, kinematics.}
}
\titlerunning{UGC\,7388}
\maketitle

\section{Introduction}

The more detailed the study of a particular galaxy is, the larger the number of
formations and features of various kinds indicative of its preceding evolution are 
detected in its structure. These formations are discovered both
in the central regions of galaxies (chemically and kinematically decoupled 
nuclei and circumnuclear structures; see, e.g., the review by Silchenko 2007)
and in their outer regions (tidal structures, stellar and gaseous disk 
warps, star streams in galactic halos, etc. -- see, e.g., Ibata et al. 2007; 
Martinez-Delgado et al. 2008).

Generally, the outer structures have a very low
surface brightness and a relatively old age, $\geq 10^8$ yr.
Their presence is usually attributed to the interactions
and mergers of galaxies as well as to the absorption
of relatively low-mass companions. The statistics of
such relics of the hierarchical galaxy formation is an
important test for the evolution models of extragalactic
objects, while their characteristics (shapes, sizes,
luminosities, etc.) can provide valuable information
about the structure of dark halos and the star formation
under unusual conditions (see, e.g., Dubinski
et al. 1996; Reshetnikov and Sotnikova 2001).

In this paper, we present the results of our morphological
and spectroscopic studies of the spiral
galaxy UGC\,7388. This galaxy has not been studied
in detail previously, although a faint extended optical
structure similar to those observed in several polar-ring
galaxies has long been pointed out in its structure
(Schweizer et al. 1983; Whitmore et al. 1990).

\section{Observations}

The observations of UGC\,7388 were performed on
April 8, 2005, with the 8.1-m Gemini North telescope
(Hawaii, USA)\footnote{Request ID GN-2005A-Q-66.}. 
We used the GMOS spectrograph in long-slit 
mode\footnote{A description of the instrument can be found at
http://www.gemini.edu/sciops/instruments/gmos/gmos-Index.html}
with the Red1 detector. The seeing
during the observations was about 1.$''$4 (FWHM).
During the observations, we obtained two optical
images of the galaxy (the $r$-G0303 filter with an effective
wavelength of 630 nm), each with an exposure
time of 5 min. These images (with a scale of 0.$''$1454
per pixel) were used to analyze the morphology of the
galaxy and to accurately set the slit for the subsequent
spectroscopic observations.

The spectroscopic observations (with the
R831+G5302 grating centered at 675 nm) were
performed at two spectrograph slit positions, P.A.=49$^{\rm o}$
(the major axis of the galaxy's main body) and
P.A.=109$^{\rm o}$ (the major axis of the brighter inclined loop) --
see Fig.\,1. The spectral range was 5700--7800\,\AA, the spectral
resolution was about 3000, and the scale of the spectra was
0.68\,\AA/pixel. At each position of the slit (its width and length 
were 0.$''$75 and 1.$'$8, respectively), we took three spectra, each 
with an exposure time of 15 min. Reproductions of the 2D
spectra for the galaxy and the spectrum of its nucleus
are shown in Fig.\,2.

We reduced the data in a standard way using the IRAF and MIDAS packages.

\begin{figure*}[!ht]
\centerline{\psfig{file=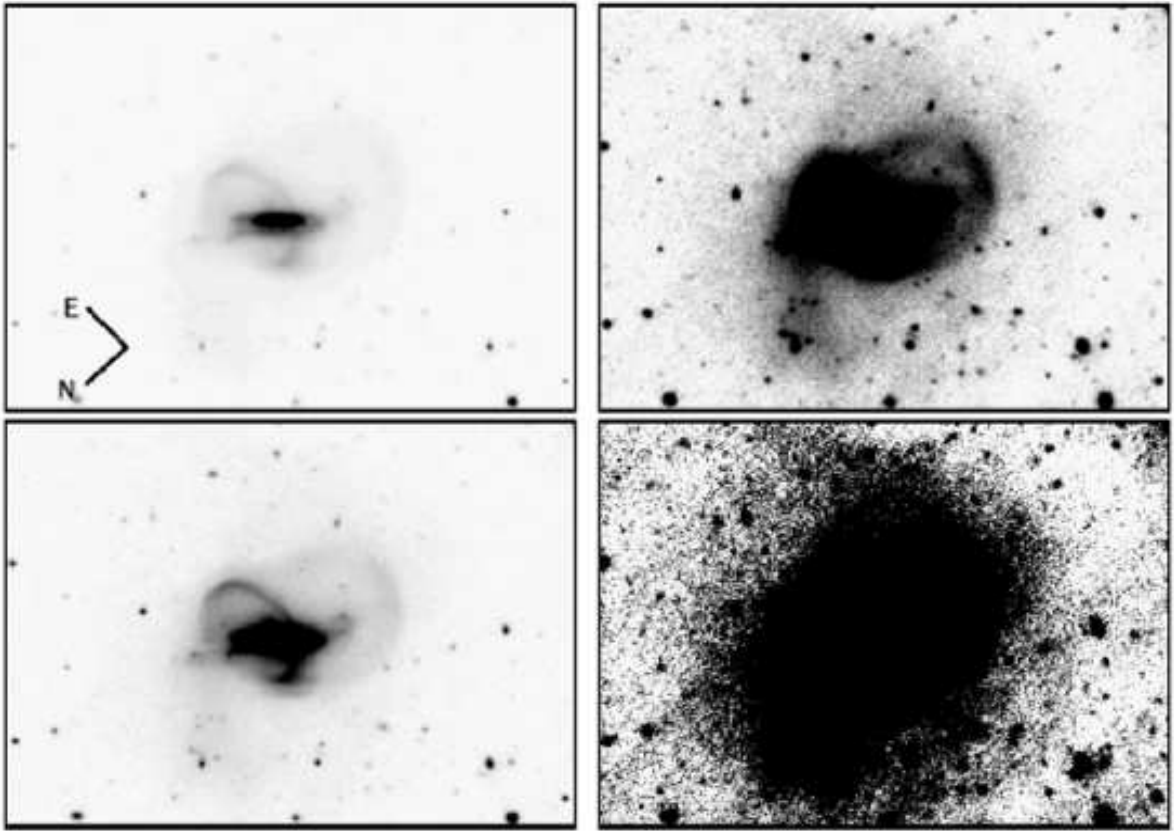,width=16cm,clip=}}
\caption{Reproductions of the image for UGC\,7388 in the $r$-G0303 filter 
(effective wavelength 630 nm) with different contrasts 
to highlight the main morphological features of the galaxy. Each
frame is $2.'4 \times 3.'3$ (62 $\times$ 86 kpc) in size.}
\end{figure*}

\section{Results}

\subsection{The morphological structure of UGC\,7388}

Figure\,1 shows reproductions of the image for
UGC\,7388 at 630 nm. In the upper left panel, the
object appears as an ordinary spiral galaxy seen at
a large angle to the line of sight (nearly edge-on).
A dust lane displaced by $2'' - 3''$ from the center
stretches along the entire NW edge of the galaxy.
A straight structure extending beyond the NE edge
of the galactic disk lies to the NW of this lane. The
SW edge of the UGC\,7388 disk is warped southward.
The disk inclination was roughly estimated from its
observed flattening ($b/a \approx 0.17$ from the SE half of
the main body undistorted by absorption) and from
the dust lane visibility conditions to be $i \geq 80^{\rm o}$.

The extended ringlike structure is slightly displaced
to the NE of the galactic center and crosses
the galactic plane at an angle of $\sim 55^{\rm o}$ (the major axis
of the ring is 47$''$ or 20 kpc; its apparent flattening is
$b/a \approx 0.6$). Previously, the presence of this feature
allowed UGC\,7388 to be considered as a probable
candidate for polar-ring galaxies (Whitmore et al. 1990).

As the image contrast increases, the ring-like
structure becomes increasingly distinct (the lower
left frame in Fig.\,1). However, the second, fainter
and more extended, ring elongated from the north
to the south also shows up (see also the upper right
frame). This ring crosses the galactic plane at an
angle of about 150$^{\rm o}$ (so that the ring-like structures
are almost orthogonal to each other); its major axis is
78$''$ (34 kpc) and its apparent flattening is $b/a \approx 0.7$

A faint diffuse feature stretching to the NW from
the NE edge of the galactic disk can be seen in the
upper right frame of Fig.\,1. This feature and the two
loops described above probably form a single continuous
structure wound around the edge-on galactic
disk. The galaxy as a whole is embedded in a faint,
almost circular (in projection) envelope $\sim$60 kpc in diameter.

Many of the features of UGC\,7388 described
above are clearly traceable on the contour map shown
in Fig.\,3.

\begin{figure}
\centerline{\psfig{file=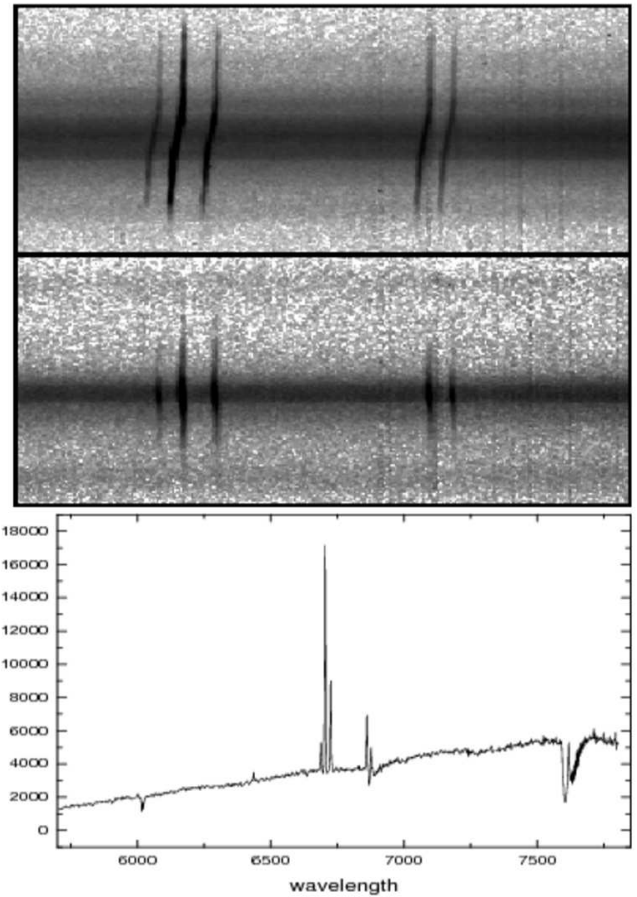,width=8.8cm,clip=}}
\caption{Reproductions of the 2D spectra for UGC\,7388 on a logarithmic 
scale (the region with the H$\alpha$, [NII], and [SII] emission 
lines is shown): the spectra with the position angles of the spectrograph 
slit (a) P.A.=49$^{\rm o}$ (NE is upward, SW is downward) and 
(b) P.A.=109$^{\rm o}$ (E is upward,W is downward). Each panel is
390\,\AA\,$\times$\,50$''$ in size; (c) the spectrum of the galactic nucleus
(in relative intensities).}
\end{figure}

\begin{figure*}
\centerline{\psfig{file=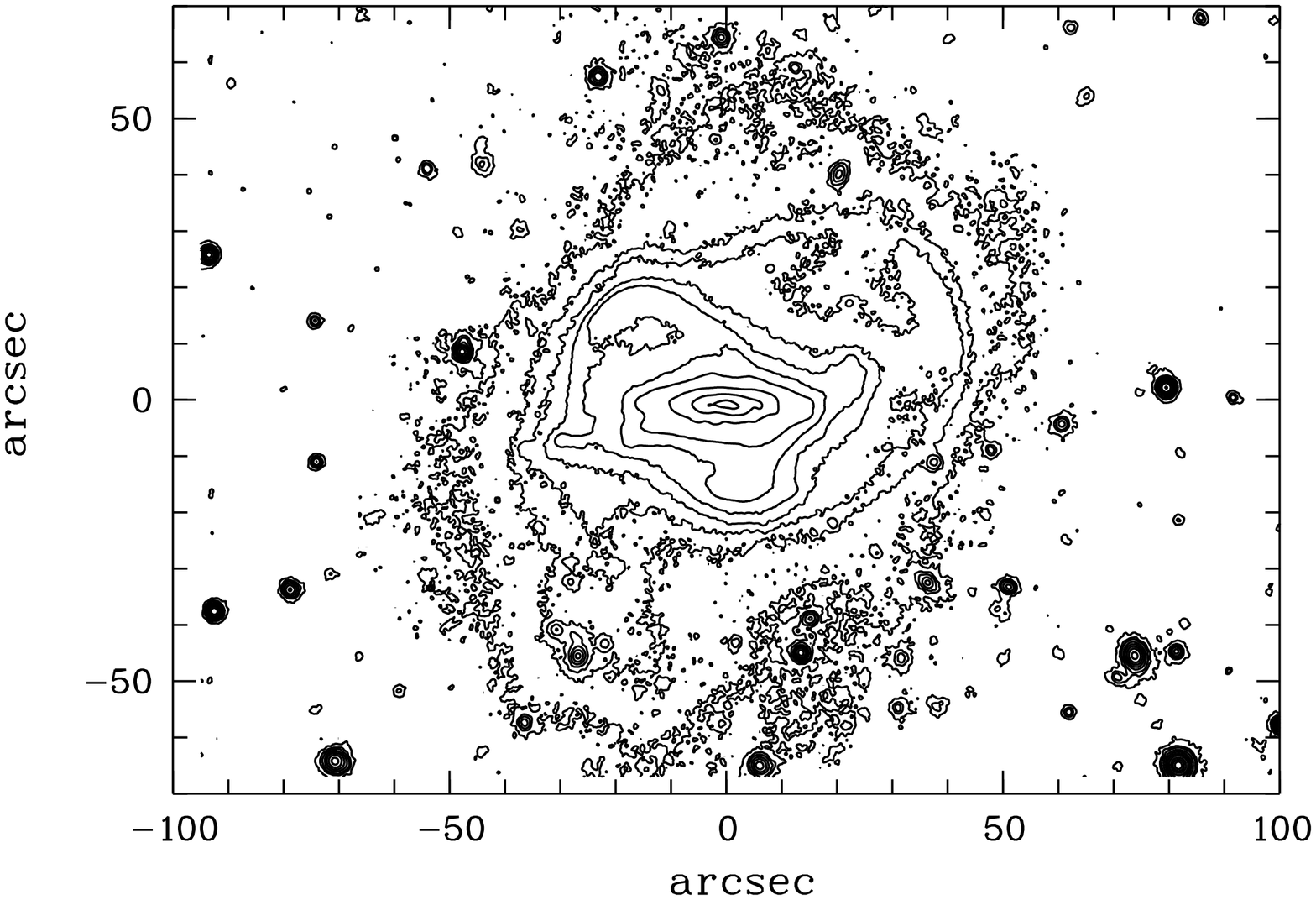,width=16cm,clip=,angle=-90}}
\caption{Isophotal map of UGC\,7388 in the $r$-G0303 filter. A surface 
brightness $\mu_R \approx 25.^m5/\Box''$ corresponds to the faintest
isophote; the subsequent isophotes are 24.6, 24.2, 23.6, 23.1, 22.6, 
22.0, 21.1, 20.4, and 19.6.}
\end{figure*}

\subsection{Photometric characteristics of UGC\,7388}

The main characteristics of the galaxy that we
found and those taken from the literature are summarized
in the table. To estimate the apparent magnitude
of UGC\,7388 in the $B$ band\footnote{All magnitudes are given 
here in the AB system of magnitudes.} and its $B-V$ color 
index, we used photometric data from the SDSS survey
(Adelman-McCarthy et al. 2007) recalculated to the $B$ and $V$ bands
(Blanton and Roweis 2007). Note that the estimated apparent magnitude 
of the galaxy, $B=15.25$, is brighter than that in the
LEDA\footnote{Lyon--Meudon Extragalactic Database.} 
($16.18 \pm 0.5$) by almost one magnitude.

The total luminosity of the loop system and the
outer envelope is comparable to the observed luminosity
of the central galaxy (within the faintest
isophote shown in Fig.\,3, the observed ratio of the
luminosity of the outer structures to the luminosity
of the central galaxy reaches $\sim$0.8). However, the
central galaxy is seen nearly edge-on and its luminosity
should be corrected for internal extinction. Following
the recommendations by Tully et al. (1998),
we estimated that the central galaxy seen face-on
should be brighter by $\sim1^m$ (in the $R$ filter).
Thus, the relative luminosity of the outer structures decreases
to $\sim 20 - 30\%$ of the luminosity of the central galaxy.

Using an approximate calibration of our frames
(based on the SDSS apparent magnitude of UGC\,7388), we found 
that the brightest parts of the smaller ring have a surface 
brightness from $\mu_R \approx 22.9^m/\Box''$ (the E and SE
regions of the ring) to $\mu_R \approx 22.3^m/\Box''$ (the W edge).
Below, the surface brightnesses are given without their units of 
measurement. The southern region of the more extended
loop exhibits a lower brightness, $\mu_R \approx 24.0 - 24.4$.
The corresponding $B$-band brightnesses (for $B-R$ color indices
of the loops from +1.0 to +1.5) are $\mu_B \approx 23.3 - 24.4$
and $\mu_B \approx 25.0 - 25.9$ for the smaller and larger rings.

\begin{table*}

%\begin{center}

\caption[]{Main characteristics of UGC\,7388}

\vspace{0.4cm}

\begin{tabular}{|l|l|l|}
\hline
Morphological type            & Sc:         &          \\
Apparent magnitude, $B$ & 15.25       &          \\
$B-V$                          & +0.8        &    \\
Diameter ($\mu_R=25.5$)         & 120\arcsec\,\, (52 kpc) &  \\
Extinction in our Galaxy, $A_B$  & 0.05          & NED$^*$ \\
Heliocentric radial velocity  & 6442 $\pm$ 11 km/s & \\
Photometric distance     & 93.8 Mpc       & NED \\
Scale                        & 0.435 kpc/1$\arcsec$ & NED \\
                               &                      &     \\
Absolute magnitude  &    &   \\
of central galaxy, $M_B^0$ & --20.4:  &   \\
Radial scale length of brightness distribution, $h$ & 6.$\arcsec$9 (3.0 kpc)&\\
Maximum rotation velocity, V$_{max}$  &   220$\pm$20 km/s    & \\ 
                                            &                      & \\
Observed HI profile width, W$_{20}$   & 471 km/s & O'Neil et al. (2004)\\
M(HI)         & 6.3 $\times$ 10$^9$\,M$_{\odot}$ & O'Neil et al. (2004) \\    
$L_{FIR}$     & 1.4 $\times$ 10$^{10}$\,$L_{\odot}$  &  NED \\
SFR$_{FIR}$   & 7 M$_{\odot}$/yr                    &      \\
\hline 
\end{tabular} \\
                            
$^*$ -- NASA/IPAC Extragalactic Database
                                        
%\end{center}
                                            
\end{table*}
                                                                           
The galaxy has no distinct bulge and the brightness
distribution along its major axis can be described
by an exponential disk with a radial scale length of
3.0 kpc. The color index of UGC\,7388 ($(B-V)_0 \approx +0.5...+0.6$ 
corrected for the inclination) and the relative abundance of
neutral hydrogen ($M({\rm HI})/L_B \approx 0.2-0.3$) are also 
consistent with the fact that it is a late-type galaxy.

The vertical $R$-band brightness distribution of the
galactic disk is distorted by the dust lane. The $J$-,
$H$-, and $K$-band images of UGC\,7388 taken from the
2MASS survey\footnote{http://www.ipac.caltech.edu/2mass/}
are more symmetric. In the $K$ band,
the ratio of the exponential scale lengths found along
the minor and major axes of the disk is $h_z/h \approx 0.4$.
Such a large apparent flattening may suggest that the
galactic plane is significantly inclined to the line of
sight ($i \leq 70^{\rm o}$ for the galaxy's true flattening
$\leq 0.15$), but the relatively small observed displacement of the
dust lane (see above) is apparently in conflict with this
assumption. The large observed ratio $h_z/h$ probably
suggests that the stellar disk of UGC\,7388 is intrinsically
thick.

\subsection{Results of spectroscopic observations}

The spectrum of the galactic nucleus is typical
of HII regions (Fig. 2). The systematic velocity of
the galaxy that we found (see the table) is in a good
agreement with the results of previous spectroscopic
observations (NED, the NASA/IPAC Extragalactic
Database) and radio observations (O'Neil et al. 2004).

\begin{figure}[!ht]
\centerline{\psfig{file=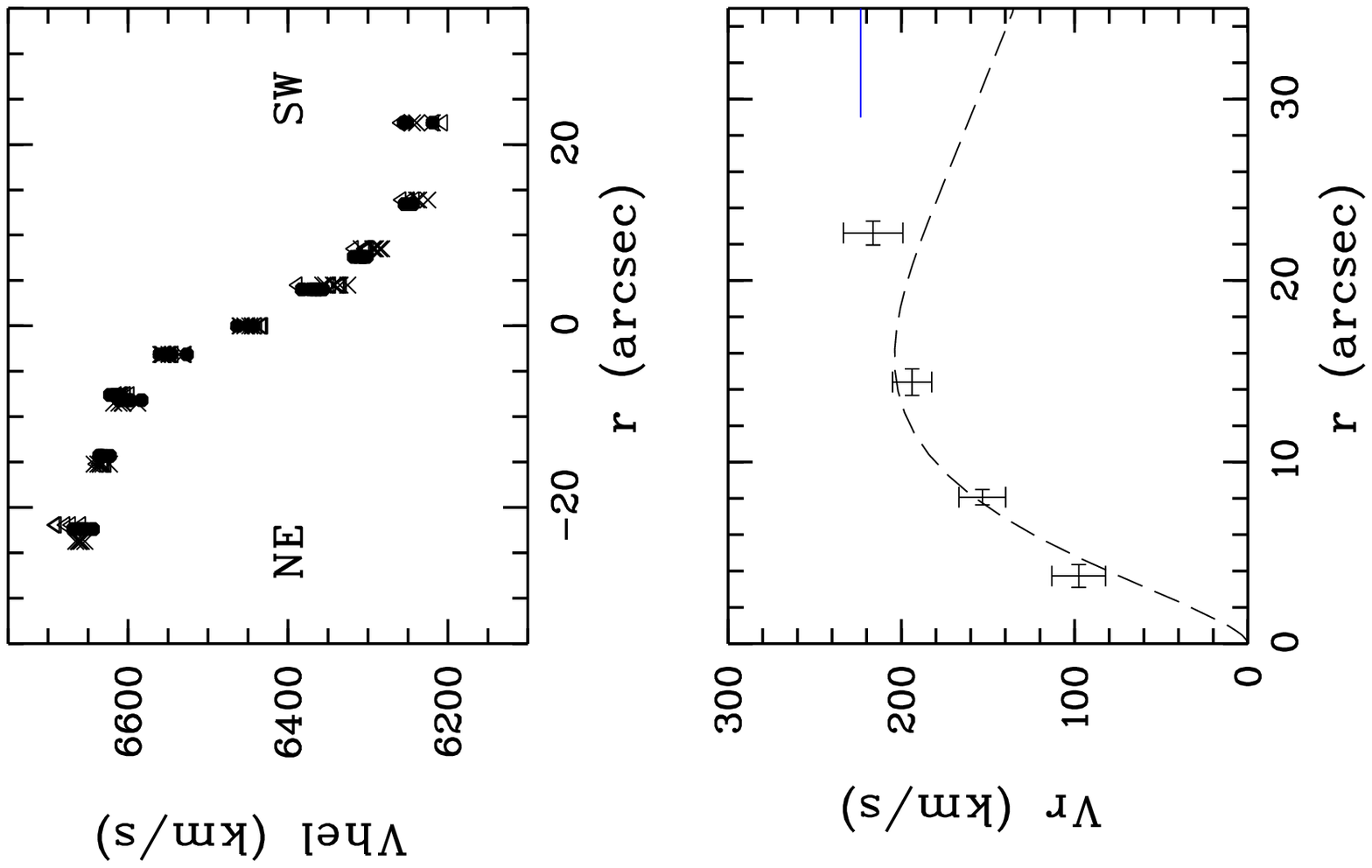,width=8.8cm,clip=,angle=-90}}
\caption{Top: Observed radial-velocity distribution along the
major axis of UGC\,7388 (different symbols indicate individual
measurements from three different spectrograms).
Bottom: Averaged rotation curve of the galaxy (crosses); the
dashed line indicates the rotation curve for an exponential
disk with a scale length of $6.''9$ and flattening $b/a=0.2$
(Monnet and Simien 1977); the horizontal bar indicates
the FWHM of the corrected HI profile.}
\end{figure}

Figure 4 presents the distribution of radial velocities
measured from the [OI]$\lambda$6300, H$\alpha$, [NII], and
[SII] emission lines along the major axis of UGC\,7388.
The errors of individual velocity measurements do not exceed 
10 km/s in the central region of the galaxy and increase
to 20--30 km/s on its periphery. The lower panel in
Fig. 4 shows the velocity distribution averaged relative to
the galaxy's dynamical center. The maximum observed rotation
velocity, $V_{max}=220$ km/s, is close to the FWHM of the
HI profile corrected in a standard way (see, e.g., Bottinelli et al. 
1990; Tully and Fouque 1985) for instrumental broadening and
turbulent gas motion (W$^{corr}_{20}$/2 = 224 km/s).

We see from Fig. 4 that the observed rotation curve within
15$''$--20$''$ of the galactic center can be
represented by an exponential disk (the figure shows
the so-called maximal disk, whose contribution to the
observed rotation curve in the central region of the
galaxy is at a maximum). However, the last point of
the rotation curve at $r=23''=10$ kpc (or, in units of the
exponential scale length, 3.3$h$), deviates from
the exponential fit, suggesting the existence of a dark
halo in which the optical disk is embedded. Within $r=23''$,
the contribution from the hidden mass is relatively small --
the ratio of the dark halo mass to the stellar disk mass is
$\sim$1/3. This estimate of the contribution from the dark halo 
is most likely a lower limit -- if the galactic disk is not 
maximal and/or if the exponential scale length of the density 
distribution in the disk is smaller than our adopted value (e.g.,
2MASS images are indicative of lower values of $h$),
then the contribution from the dark halo can be considerably 
larger.

\begin{figure}[!ht]
\centerline{\psfig{file=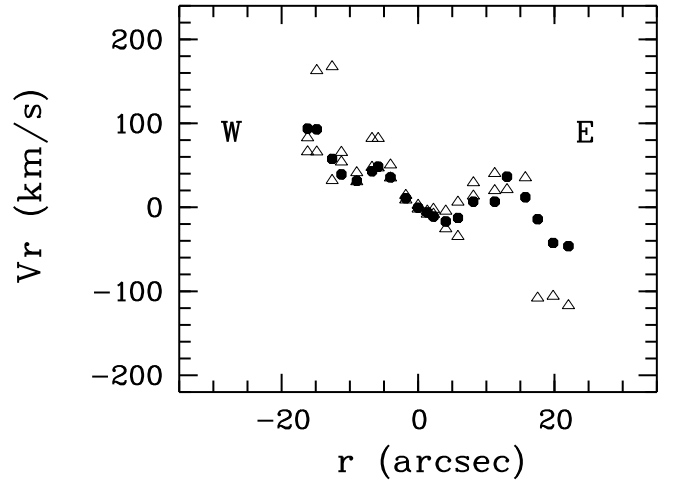,width=8.8cm,clip=,angle=-90}}
\caption{Distribution of relative radial velocities at 
P.A.=109$^{\rm o}$ (the filled circles and triangles represent the data
based on H$\alpha$ and [NII]$\lambda$6548,6583, respectively).}
\end{figure}

The radial-velocity distribution for the H$\alpha$ and [NII] lines
at P.A.=109$^{\rm o}$ is shown in Fig. 5. Obviously,
the ring-like structure (the brighter and more
compact of the two structures described above)
shows evidence of a large-scale rotation relative to
the central galaxy: the W part of the ring recedes
from us with a relative velocity of $\sim$100 km/s, while its E
part approaches us with a velocity of $\sim$50--100 km/s. 
Thus, UGC 7388 could have
been classified as a kinematically confirmed polar-ring
galaxy, but the presence of a second loop-like
structure and the noncoincidence of the loop centers
with the galactic nucleus are in conflict with the
classical definition of this type of galaxies (Whitmore
et al. 1990).

In general, the radial-velocity distribution
of the emitting gas along the major axis of the loop appears
non-monotonic: a local velocity maximum is observed
at a distance of 10$''$--15$''$ to the east of the nucleus.
This feature may be related, for example, to the interaction
between the gas subsystems of the loop
and the main galaxy. In addition, to all appearances,
the general velocity field of UGC\,7388 is a complex,
multi-component one, since it reflects the capture
and disruption of a relatively massive companion (see
the subsequent discussion).

\section{Discussion}

The data from the table and the preceding discussion
show that UGC\,7388 is a typical giant spiral
galaxy. It is similar to the Milky Way Galaxy in luminosity,
maximum rotation velocity, and exponential
scale length of the brightness distribution. The main
differences between UGC\,7388 and the Milky Way
are the absence of a prominent bulge and a
slightly higher HI abundance.

The most remarkable feature of UGC\,7388 is that
it is surrounded by two faint extended loop-like structures
(Figs. 1 and 3). The existence of the brighter
structure with $\mu_B \sim 24$ has already been known (its
presence was the reason why the galaxy was included
in the list of candidate polar-ring galaxies); the fainter
structure ($\mu_B \sim 25.5$, the diameter exceeds 30 kpc) is
described here for the first time.

One of the standard scenarios for the formation
of outer structures in galaxies is the tidal capture
of part of the matter from a closely passing galaxy.
However, UGC\,7288 is a relatively isolated object --
there are no other objects of comparable luminosity
within several hundred kpc of UGC\,7388 (Brocca
et al. 1997). Consequently, the assumption that the
capture and tidal disruption of a relatively massive
companion is observed in the case of UGC\,7388
seems most natural. Emission lines are observed in
the region of the brighter loop (Fig. 5) and, hence, this
companion most likely contains gas and is a spiral
or irregular galaxy.

An upper limit on the mass of the companion
being disrupted can be roughly estimated from the
observed luminosity of the structures outside the disk
of the main galaxy to be 20--30\% of the main galaxy's
mass. This is a gross overestimate, since the matter
of the central galaxy itself snatched away during the
interaction with the companion can also contribute
to the luminosity of the outer structures. In addition,
the observed (at least in the brighter loop, Fig. 2)
emission lines can also contribute to the observed $R$ band
luminosity of the outer structures. These effects
are fairly difficult to take into account. Therefore, as
a rough estimate, we assume that the relative mass
of the companion can reach $\sim0.2\pm0.1$ of the main galaxy's
mass.

As for the interaction geometry, the companion
most likely approached the NE edge of the
galaxy from the NW (see Fig. 1); it described initially
the larger (fainter) loop and subsequently the
smaller (brighter) one. In general, the companion's
orbit traced by the tidal structures is very similar to
the results of numerical simulations presented by Villalobos
and Helmi (2008) (see Fig. 3 in their paper).
The time it takes for the companion to be disrupted is
$\sim$1 Gyr (estimated from the total length of the loops
and the velocity of the companion, 200 km/s).

What suggests that a prolonged interaction of
two galaxies is observed in the case of UGC\,7388?
First of all, it is the fact that the main galaxy has
a perturbed structure. For example, a linear structure,
which may be a tidal tail seen nearly edge-on,
stretches along its NW edge. The SW edge of the disk
is strongly warped (the warp of the opposite part of the
disk is probably masked by the previously mentioned
straight structure and the dust lane). The existence
of such deformations of the stellar disks in galaxies
is often related to their interactions and the presence
of close companions (Reshetnikov and Combes 1998,
1999). There is evidence of an excess in thickness of the
stellar disk in UGC\,7388 (see above), which is also a
standard consequence of the gravitational interaction
between galaxies (Reshetnikov and Combes 1997).
Additional signatures of the absorption of a companion
include the presence of an extended, nearly spherical
envelope $\sim$60 kpc in diameter surrounding
the galaxy and the relatively high infrared luminosity
of UGC\,7388 (see the table).

Very few objects similar to UGC\,7388 (spiral galaxies 
surrounded by extended loop-like
structures of low surface brightness visible in the
optical range) are known to date. As the closest
example, we can mention the Milky Way Galaxy,
whose disk is crossed by an extended star stream
produced by the dwarf galaxy Sgr I being disrupted
(Ibata et al. 1995). Another example is the nearly
edge-on galaxy NGC\,5907, whose main body is surrounded
by a very faint ($\mu_R \approx 27-28$) loop-like structure
$\sim$40 kpc in diameter (Shang et al. 1998). The
observed morphology of this loop is well described
as the remnant of a disrupted low-mass ($\sim10^{-3}$ of the 
main galaxy's mass) companion (Reshetnikov and Sotnikova 
2000). A large-scale system of faint ($\mu_R \approx 27$)
tidal loops has recently been discovered
around the spiral galaxy NGC\,4013, which is also
seen nearly edge-on (Martinez-Delgado et al. 2008).

The three galaxies (NGC\,5907, NGC\,4013, and
UGC\,7388) share several common features. First,
they are all seen at a large angle to the line of sight,
edge-on. Such a preferential orientation is undoubtedly
the result of observational selection, since the
outer structures in galaxies with a smaller inclination
to the line of sight will be projected onto their
main bodies and will remain undetectable. Second,
all three galaxies are relatively isolated objects. This
peculiarity probably stems from the fact that the presence
of nearby galaxies can disrupt the thin, ``coherent''
outer structures. Third, all three galaxies exhibit
large-scale warps of their stellar and/or gaseous
disks. As was noted above, such deformations of the
galactic disks can be caused by their gravitational
interaction with the companions being absorbed.

An important difference between UGC\,7388 and
other similar objects is that the disruption of not a
dwarf, but of a relatively massive companion is observed in
this galaxy. Accordingly, the tidal loops have a rather
high surface brightness, which allows their geometry
and even kinematics to be investigated without any
special tricks. This turns UGC\,7388 into a unique
laboratory for studying such topical questions as the
disruption of a companion in the gravitation field of a
massive galaxy and the influence of a companion on
the structure of the main galaxy (vertical heating of
the stellar disk, the formation of its large-scale warp,
triggered star formation, etc.). Realistic numerical
simulations of UGC\,7388 could allow the extent,
mass, and, possibly, shape of the dark halo in this
galaxy to be estimated.

\bigskip
\section*{Acknowledgments}
This work is based on the observational data obtained
at the Gemini Observatory, which is operated
by the Association of Universities for Research
in Astronomy (AURA) under a cooperative agreement
with NSF on behalf of the Gemini Union: NSF
(USA), STFC (United Kingdom), NRC (Canada),
CONICYT (Chile), ARC (Australia), CNPq (Brazil),
and SECYT (Argentina). The observations were performed
under ID GN-2005A-Q-66. The study was
supported by the Russian Foundation for Basic Research
(project no. 06-02-16459).

\section*{REFERENCES}

\indent

1.\,J.K.\,Adelman-McCarthy, M.A.\,Agueros, S.S.\,Allam,
et al., Astrophys. J. Suppl. Ser. 172, 634 (2007).

2.\,M.R.\,Blanton and S. Roweis, Astron. J. 133, 734
(2007).

3.\,L.\,Bottinelli, L.\,Gouguenheim, P.\,Fouque, and G.\,Paturel,
Astron. Astrophys. Suppl. Ser. 82, 391 (1990).

4.\,Ch.\,Brocca, D.\,Bettoni, and G.\,Galletta, Astron. Astrophys.
326, 907 (1997).

5.\,J.\,Dubinski, J.\,Mihos, and L.\,Hernquist, Astrophys.
J. 462, 576 (1996).

6.\,R.A.\,Ibata, G.\,Gilmore, and M.J.\,Irwin, Mon. Not.
R. Astron. Soc. 277, 781 (1995).

7.\,R.\,Ibata, N.F.\,Martin, M.\,Irwin, et al., Astrophys. J.
671, 1591 (2007).

8.\,D. Martinez-Delgado, M.\,Pohlen, R.J.\,Gabany, et al.,
arXiv:0801.4657v1 (2008).

9.\,G.\,Monnet and F.\,Simien, Astron. Astrophys. 56, 173
(1977).

10.\,K.\,O'Neil, G.\,Bothun, W.\,van Driel, and D.\,Monnier
Ragaigne, Astron. Astrophys. 428, 823 (2004).

11.\,V.\,Reshetnikov and F.\,Combes, Astron. Astrophys.
324, 80 (1997).

12.\,V.\,Reshetnikov and F.\,Combes, Astron. Astrophys.
337, 9 (1998).

13.\,V.\,Reshetnikov and F.\,Combes, Astron. Astrophys.
Suppl. Ser. 138, 101 (1999).

14.\,V.P.\,Reshetnikov and N.Ya.\,Sotnikova, Pis'ma
Astron. Zh. 26, 333 (2000) [Astron. Lett. 26, 277 (2000)].

15.\,V.P.\,Reshetnikov and N.Ya.\,Sotnikova, Astron. Astrophys.
Trans. 20, 111 (2001).

16.\,F.\,Schweizer, B.C.\,Whimore, and V.C.\,Rubin, Astron.
J. 88, 909 (1983).

17.\,Zh.\,Shang, Zh.\,Zheng, E.\,Brinks, et al., Astrophys. J.
504, 23L (1998).

18.\,O.K.\,Silchenko, Astronomy: Traditions, Present,
Future Ed. by V.V.\,Orlov, V.P.\,Reshetnikov, and
N.Ya.\,Sotnikova (St.-Petersburg State University,
St. Petersburg, 2007), p. 63.

19.\,R.B.\,Tully and P.\,Fouque, Astrophys. J. Suppl. Ser.
58, 67 (1985).

20.\,R.B.\,Tully, M.J.\,Pierce, J.-Sh.\,Huang, et al., Astron.
J. 115, 2264 (1998).

21.\,A.\,Villalobos and A.\,Helmi, arXiv:0803.2323v1 (2008).

22.\,B.C.\,Whitmore, R.A.\,Lucas, D.B.\,McElroy, et al.,
Astron. J. 100, 1489 (1990).

\end{document}